\documentstyle[12pt]{article}

\begin{document}
\title{On ortho-positronium and gauge}
\author{G.V.Efimov  \vspace*{0.2\baselineskip}\\
 \itshape Bogoliubov Laboratory of Theoretical Physics,\\
 \itshape Joint Institute for Nuclear Research,\\
{\it 141980 Dubna, Russia}\vspace*{0.2\baselineskip} }
%
% \date{Pacs Numbers: 13.20.-v, 13.20.Fc, 13.20.He, 24.85.+p}
%
\maketitle
%..........

\begin{abstract}
Binding energy of the $1^-$ state (ortho-positronium) in QED is
calculated using the one-photon exchange Bethe-Salpeter equation in
the Feynman and Coulomb gauges for different coupling constants
$\alpha$. Calculations show there is a remarkable difference in
values of the binding energy for different coupling constants in
these two gauges.
\end{abstract}

\section{Introduction}

The purpose of this paper is to calculate binding energy of the
two-fermion $1^-$ system (ortho-positronium) in quantum
electrodynamics (QED) using the Bethe-Salpeter (BS) equation for the
Feynman and Coulomb gauges for different values of the coupling
constants $\alpha$ by a variational method suggested recently
\cite{Ef-frm} and see the difference in values of the binding energy
for different coupling constants in these two gauges.

The motivation for this study is to look for an acceptable method of
investigation of bound states in systems like quantum chromodynamics
(QCD) in which the coupling constant $\alpha_s\sim0.1\div0.5$ is not
too small as in the QED $\alpha={1\over137}=0.0073$. In general, it
is generally agreed that the BS equation in the one-photon or
one-gluon approximation appears as a quite acceptable instrument for
study of a bound state problem if the coupling constant is small
enough. In principle, it is hoped that this approximation gives the
main contribution to binding energy. However, the BS equation in
gauge theories is not gauge invariant in this approximation and in
QCD the coupling constant is not so small as in QED. Therefore, we
should know what difference in binding energies arises for different
gauges and different coupling constants.

Most computations of electromagnetic bound states use the Coulomb
gauge and the Breit potential with relativistic corrections (see,
for example, \cite{Berest,Eid,Grein,Beck}). This approach is most
compatible with experimental data.

Long time ago an attempt was made \cite{Love} to consider gauge
properties of the BS-equation for the two-fermion electromagnetic
bound state for different covariant and axial gauges. Naturally, it
was found that different gauges give different results in the
highest orders in electromagnetic coupling constant $\alpha$.

\section{Lagrangian.}

We perform all calculations in the Euclidean metrics. The QED
lagrangian  describing electrons and photons looks like
\begin{eqnarray}\label{L}
&& L=-{1\over4}F_{\mu\nu}^2(x)+
(\overline{\psi}(x)(\hat{p}+e\hat{A}(x)-m)\psi(x)),\\
&& F_{\mu\nu}(x)=\partial_\mu A_\nu(x)-\partial_\nu
A_\mu(x).\nonumber
\end{eqnarray}

The electron propagator has the standard form
\begin{eqnarray}\label{S}
&& S(x-x')=\left\langle{\rm T}\left[
\Psi(x)\overline{\Psi}(x')\right]\right\rangle =
\int{dp\over(2\pi)^4}{e^{ip(x-x')}\over m-i\hat{p}}
\end{eqnarray}

The photon propagator is defined by the formula
\begin{eqnarray}\label{D}
&& D_{\mu\nu}(x)=\langle{\rm T}[A_\mu(x)A_\nu(0)]\rangle=
\int{dk\over(2\pi)^4}\tilde{D}_{\mu\nu}(k)e^{ikx}
\end{eqnarray}
where in the  Feynman gauge
\begin{eqnarray}
\label{DF} && \tilde{D}_{\mu\nu}(k)={\delta_{\mu\nu}\over
k^2},~~~~~~D_{\mu\nu}(y)=\delta_{\mu\nu}D(y),~~~~~D(y)={1\over(2\pi)^2y^2}.
\end{eqnarray}
and in the Coulomb gauge
\begin{eqnarray}
\label{DC} && \tilde{D}_{\mu\nu}(k)=\left\{\begin{array}{l}
\left[\delta_{ij}-{k_i k_j\over{\bf k}^2}\right]{1\over k^2}\\
\\
-{1\over{\bf k}^2}\\
\end{array}\right.
\end{eqnarray}

\subsection{States of electron-positron system}

Let $S$ be spin and $L$ be orbital~momenta. Then the total parity of
electron-positron system is defined as $P=(-1)^{1+L}$. For
para-positronium $S=0,~J^P=0^-$ and for ortho-positronium
$S=1,~J^P=1^-$.

Quantum numbers of electron-positron currents
\begin{eqnarray}
\label{OJ} \left(\overline{\Psi}(-{\bf k})O_J\Psi({\bf k})\right)
\sim \left(\overline{v}(-{\bf k})O_J u({\bf k})\right) b(-{\bf
k})a({\bf k})
\end{eqnarray}
can be determined in the nonrelativistic representations of electron
and positron wave functions
\begin{eqnarray*}
&& \Psi({\bf k})\sim u({\bf k})a({\bf k})+v({\bf k})b^+({\bf k}),\\
&& \overline{\Psi}({\bf k})\sim \overline{u}({\bf k})a^+({\bf
k})+\overline{v}({\bf k})b({\bf k}),
\end{eqnarray*}
where
\begin{eqnarray*}
&& u({\bf k})=\left(\begin{array}{c}
1 \\
{\mbox{\boldmath$\sigma k$}\over 2m}\\
\end{array}\right)\chi,~~~~~
v({\bf k})=\left(\begin{array}{c}
{\mbox{\boldmath$\sigma k$}\over 2m}\\
1 \\
\end{array}\right)\chi,\\
&& \overline{v}(-{\bf k})= \chi^+\left(\begin{array}{cc}
-{\mbox{\boldmath$\sigma k$}\over 2m}, & -1 \\
\end{array}\right)
\end{eqnarray*}
Quantum numbers of all possible electron-positron currents are
represented in Table 1.
\begin{center}

\newpage

Table 1. Quantum numbers of relativistic currents

\vspace{.5cm}

\begin{tabular}{|c|c|c|c|c|c|c|}
\hline
&&&&&&\\
$J$ &  $(\overline{\Psi}O_J\Psi)$   & $S$ & $L$  & $J$ &
$P=(-1)^{1+L}$ & $J^P$ \\
&&&&&&\\
\hline
&&&&&&\\
$S$ &  $(\overline{\Psi}\Psi)~\Longrightarrow~
(\mbox{\boldmath$\sigma k$})$   & $1$ & $1$  & $0$ &$+1$ & $0^+$ \\
&&&&&&\\
\hline
&&&&&&\\
& $(\overline{\Psi}\gamma_0\Psi)~\Longrightarrow~
(\mbox{\boldmath$\sigma k$})$   & $1$ & $1$  & $0$ & $+1$&$0^+$\\
$V$&&&&&&\\
& $(\overline{\Psi}\mbox{\boldmath$\gamma$}\Psi)~\Longrightarrow~
\mbox{\boldmath$\sigma$}$   & $1$ & $0$  & $1$ & $-1$&${\bf 1^-}$\\
&&&&&&\\
\hline
&&&&&&\\
&$(\overline{\Psi}i\gamma_0\mbox{\boldmath$\gamma$}\Psi)~\Longrightarrow~
\mbox{\boldmath$\sigma$}
$   & $1$ & $0$  & $1$ & $-1$&${\bf 1^-}$ \\
$T$&&&&&&\\
&$(\overline{\Psi}\sigma_{ij}\Psi)~\Longrightarrow~
[\mbox{\boldmath$\sigma\times k$}]$ & $1$ & $1$  & $1$ & $+1$&$1^+$ \\
&&&&&&\\
\hline
&&&&&&\\
& $(\overline{\Psi}\gamma_5\gamma_0\Psi)~\Longrightarrow~1
$   & $0$ & $0$  & $0$ & $-1$&${\bf 0^-}$ \\
$A$&&&&&&\\
&$(\overline{\Psi}\gamma_5\mbox{\boldmath$\gamma$}\Psi)~\Longrightarrow~
[\mbox{\boldmath$\sigma\times k$}]$   & $1$ & $1$  & $1$ & $+1$&$1^+$ \\
&&&&&&\\
\hline
&&&&&&\\
$P$ &  $(\overline{\Psi}\gamma_5\Psi)~\Longrightarrow~1$
& $0$ & $0$  & $0$ & $-1$&${\bf 0^-}$ \\
&&&&&&\\
\hline
\end{tabular}
\end{center}

One can see that only vector and tensor currents have the
ortho-positronium quantum numbers. In other words the
ortho-positronium should be described by a mixture of vector $V$ and
tensor $T$ relativistic currents.

\subsection{One photon exchange and ortho-positronium currents.}

In order to extract the currents having ortho-positronium quantum
numbers in the one photon exchange approximation it is convenient to
use the method of generating functional. We have
\begin{eqnarray}\label{Z}
Z&=&\int\!\!\!\int D\Psi D\overline{\Psi}DA~
e^{(\overline{\Psi}S^{-1}\Psi)- {1\over2}(A_\mu
D_{\mu\nu}^{-1}A_\nu)- e((\overline{\Psi}\gamma_\mu\Psi)A_\mu)}
=\int\!\!\!\int D\Psi D\overline{\Psi}
e^{(\overline{\Psi}S^{-1}\Psi)+W_2},\nonumber\\
\end{eqnarray}
where
\begin{eqnarray}\label{W2}
W_2&=&{e^2\over2}\int\!\!\!\int dx_1dx_2
(\overline{\Psi}(x_1)\gamma_\mu\Psi(x_1))
D_{\mu\nu}(x_1-x_2)(\overline{\Psi}(x_2)\gamma_\nu\Psi(x_2))\\
&=&-{e^2\over2}\int\!\!\!\int dx_1dx_2
(\overline{\Psi}_\alpha(x_1)\Psi_\sigma(x_2))(\overline{\Psi}_\rho(x_2)\Psi_\beta(x_1))
(\gamma_\mu)_{\alpha\beta}D_{\mu\nu}(x_1-x_2)(\gamma_\nu)_{\rho\sigma}.\nonumber
\end{eqnarray}
Let us introduce the new variables
\begin{eqnarray*}
&& x_1=x+{y\over2},~~~x_2=x-{y\over2}
\end{eqnarray*}

The Firz transformations should be used to extract the vector and
tensor currents
\begin{eqnarray*}
&&(\gamma_\mu)_{\alpha\beta}D_{\mu\nu}(y)(\gamma_\nu)_{\rho\sigma}\\
&&=\sum\limits_{J_1,J_2}(O_{J_1})_{\alpha\sigma}(O_{J_2})_{\rho\beta}
{1\over16}{\rm Tr}[O_{J_1}i\gamma_\mu
O_{J_2}i\gamma_\nu]D_{\mu\nu}(y)\\
&&\Longrightarrow-(\gamma_j)_{\alpha\sigma}{\cal
D}^V_{jj'}(y)(\gamma_{j'})_{\rho\beta}-
(i\gamma_0\gamma_j)_{\alpha\sigma}{\cal
D}^T_{jj'}(y)(i\gamma_0\gamma_{j'})_{\rho\beta} .
\end{eqnarray*}

\section{Feynman gauge}

In the Feynman gauge (\ref{DF}) we have
\begin{eqnarray}\label{DFVT}
&&{\cal D}^V_{ij}(y)={\delta_{ij}\over2}D(y),~~~~~~~{\cal
D}^T_{jj'}(y)=0.
\end{eqnarray}
Thus, in the Feynmam gauge the ortho-positronium is described by the
vector current only.

For the vector-vector part of the one-photon exchange contribution
one can get
\begin{eqnarray*}
W_2&\Longrightarrow&{e^2\over2}\int\!\!\!\int dxdy
J_i(x,y){\cal D}_{ij}(y)J_j(x,-y)\\
&=&{e^2\over4}\int\!\!\!\int dxdy J_j(x,y) D(y)J_j(x,-y),\\
J_j(x,y)&=&\left(\overline{\Psi}\left(x+{y\over2}\right)
\gamma_j\Psi\left(x-{y\over2}\right)\right)
=\left(\overline{\Psi}\left(x\right)e^{i{y\over2}\stackrel{\leftrightarrow}{p}_x}
\gamma_j\Psi\left(x\right)\right)\\
&&\stackrel{\leftrightarrow}{p}_x= {1\over
i}\left(\stackrel{\leftarrow}{\partial}_x-
\stackrel{\rightarrow}{\partial}_x\right)
\end{eqnarray*}
Let us introduce an orthonormal system of functions
\begin{eqnarray}\label{UQ}
&& \{U_Q(y)\}=\{U_{n\kappa lm}(y)\}=\left\{
\begin{array}{l}
\int dy~U_{Q_1}^*(y)U_{Q_2}(y)=\delta_{Q_1,Q_2},\\
\\
\sum\limits_Q~U_Q(y_1)U_Q^*(y_2)=\delta(y_1-y_2),\\
\end{array}\right.
\end{eqnarray}
As long as for the ortho-positronium state $Q=0$ we restrict
ourselves to the function  $U_0(y)=U(y)$ with normalization
$(UU)=1$.

Let us perform the following transformations:
\begin{eqnarray}\label{Current}
W_2&=&{e^2\over4}\int dx\int\!\!\!\int dy_1dy_2
J_i(x,y_1)\sqrt{D(y_1)}
\delta(y_1-y_2)\sqrt{D(y_2)}J_j(x,-y_2)\nonumber\\
&\Longrightarrow&{e^2\over4}\int dx~{\cal J}_j(x){\cal
J}_j(x),\nonumber\\
{\cal J}_j(x)&=&\int dy~\sqrt{D(y)} U(y)J_i(x,y)=
\left(\overline{\Psi}(x)
V(\stackrel{\leftrightarrow}{p}_x)\gamma_j\Psi(x)\right),
\end{eqnarray}
where the vertex is
\begin{eqnarray}\label{Vertex}
&&V(\stackrel{\leftrightarrow}{p}_x)=\int dy~
\sqrt{D(y)}U(y)e^{i{y\over2}\stackrel{\leftrightarrow}{p}_x},
\end{eqnarray}
The generating functional $Z$ containing the ortho-positronium
vector current can be transformed as
\begin{eqnarray}
\label{Zorth}
Z&=&\int\!\!\! D\Psi D\overline{\Psi}
e^{(\overline{\Psi}S^{-1}\Psi)+{e^2\over4}\int dx~ {\cal J}_j(x){\cal J}_j(x)}
\nonumber\\
&=&\int D{\cal B} e^{-{1\over2}({\cal B}_j{\cal B}_j)}
\int\!\!\!\int D\Psi D\overline{\Psi}
e^{(\overline{\Psi}S^{-1}\Psi)+{e\over\sqrt{2}}({\cal B}_j{\cal J}_j)}\nonumber\\
&=&\int D{\cal B} e^{-{1\over2}({\cal B}_j{\cal B}_j)}\int\!\!\!\int
D\Psi D\overline{\Psi}
e^{(\overline{\Psi}S^{-1}\Psi)+{e\over\sqrt{2}}
(\overline{\Psi} V\gamma_j{\cal B}_j\Psi)}\nonumber\\
&=&\int D{\cal B} e^{-{1\over2}({\cal B}_j{\cal B})+
{\rm Tr}\ln[1+{e\over\sqrt{2}}V\gamma_j{\cal B}_jS]}\nonumber\\
&=&\int D{\cal B} e^{-{1\over2}({\cal B}_j{\cal
B}_j)-{e^2\over4}{\rm
Tr}[V\gamma_i{\cal B}_iSV\gamma_j{\cal B}_jS]+O(e^4)}\nonumber\\
&=&\int D{\cal B}~ e^{-{1\over2}({\cal
B}_i[\delta_{ij}-\Pi_{ij}]{\cal B}_j)+O(e^4)}
\end{eqnarray}
with
\begin{eqnarray*}
&&-{e^2\over2}{\rm Tr}[V\gamma_i{\cal B}_iSV\gamma_j{\cal B}_jS]=
\int\!\!\!\int dx_1dx_2{\cal B}_i(x_1)\Pi_{ij}(x_1-x_2){\cal B}_j(x_2)\\
&&=\int
{dq\over(2\pi)^4}\tilde{B}_i^+(q)\tilde{\Pi}_{ij}(q)\tilde{B}_j(q)
\end{eqnarray*}

The polarization operator is defined by
\begin{eqnarray*}
&&\Pi_{ij}(x_1-x_2)=-{e^2\over2}{\rm Tr}\left[V
\left(\stackrel{\leftrightarrow}{p}_{x_1}\right)\gamma_iS(x_1-x_2)
V\left(\stackrel{\leftrightarrow}{p}_{x_2}\right)\gamma_j
S(x_2-x_1)\right]\\
&&=\int{dq\over(2\pi)^4}e^{iq(x_1-x_2)}\tilde{\Pi}_{ij}(q),
\end{eqnarray*}
where
\begin{eqnarray}
\label{Polar}
&&\tilde{\Pi}_{ij}(q)=-{e^2\over2}\int
\!{dk\over(2\pi)^4}V^2(k)~{\rm
Tr}\left[\gamma_i\tilde{S}\left(k+{q\over2}\right)
\gamma_j{\tilde S}\left(k-{q\over2}\right)\right]\nonumber\\
&&=-{e^2\over2} \int{dk\over(2\pi)^4}V^2(k) {{\rm
Tr}\left[\gamma_i\left(m+i\hat{k}+i{\hat{q}\over2}\right)\gamma_j
\left(m+i\hat{k}-i{\hat{q}\over2}\right)\right]\over
\left(m^2+\left(k+{q\over2}\right)^2\right)
\left(m^2+\left(k-{q\over2}\right)^2\right)}\nonumber\\
&&={e^2\over2}\int{dk\over(2\pi)^4}V^2(k)
{4\left[\left(m^2+k^2-{q^2\over4}\right)\delta_{ij}
-2\left(k+{q\over2}\right)_i\left(k-{q\over2}\right)_j\right]\over
\left(m^2+\left(k+{q\over2}\right)^2\right)
\left(m^2+\left(k-{q\over2}\right)^2\right)}\nonumber\\
&&\Rightarrow 2e^2\int{dk\over(2\pi)^4}V^2(k)
{\left[k^2-{2\over3}{\bf k}^2\right]
+\left[m^2+{M^2\over4}\right]\over
\left(m^2+\left(k+{q\over2}\right)^2\right)
\left(m^2+\left(k-{q\over2}\right)^2\right)}\cdot\delta_{ij}
\end{eqnarray}
We choose the frame where $q=({\bf 0},iM)$.

According to \cite{Ef-frm}, we divide the polarization operator
(\ref{Polar}) in two parts
\begin{eqnarray*}
&&\tilde{\Pi}_{ij}(q)=\left[\tilde{\Pi}_0(q)+\tilde{\Pi}_I(q)\right]\delta_{ij},
\end{eqnarray*}
where
\begin{eqnarray*}
&&\tilde{\Pi}_0(q)=2e^2\int{dk\over(2\pi)^4}V^2(k)~
{\left(k^2-{2\over3}{\bf k}^2\right)
\over\left(m^2+\left(k+{q\over2}\right)^2\right)
\left(m^2+\left(k-{q\over2}\right)^2\right)}\\
&&=2e^2\int{d{\bf k}dk_4\over(2\pi)^4}V^2({\bf k},k_4)~
{\left({1\over3}{\bf k}^2+k_4^2\right) \over \left({\bf
k}^2+k_4^2+m^2-{M^2\over4}\right)^2+M^2k_4^2},
\end{eqnarray*}
is responsible for a continuous spectrum, and
\begin{eqnarray*}
&&\tilde{\Pi}_I(q)=2e^2\int{dk\over(2\pi)^4}V^2(k)~{\left(
m^2+{M^2\over4}\right)\over
\left(m^2+\left(k+{q\over2}\right)^2\right)
\left(m^2+\left(k-{q\over2}\right)^2\right)}\\
&&=2e^2\int{d{\bf k}dk_4\over(2\pi)^4}V^2({\bf k},k_4)~
{\left(m^2+{M^2\over4}\right) \over \left({\bf
k}^2+k_4^2+m^2-{M^2\over4}\right)^2+M^2k_4^2},
\end{eqnarray*}
is responsible for a bound state.

The vertex (\ref{Vertex}) looks like
\begin{eqnarray*}
&& V(k)=V({\bf k},k_4)=\int\! d{\bf y}\int dy_4~{\Psi({\bf
y},y_4)e^{i{\bf yk}+ik_4y_4}\over(2\pi)^2({\bf y}^2+y_4^2)}
\end{eqnarray*}
with $U(y)=\sqrt{D(y)}~\Psi({\bf y},y_4)$, where $\Psi({\bf y},y_4)$
is a wave function.

Next we should extract in the kernel of (\ref{Zorth}) the part that
is responsible for the ortho-positronium bound state
\begin{eqnarray*}
&& ({\cal B}[I-\Pi]{\cal B})=({\cal B}[I-\Pi_0-\Pi_I]{\cal B})
\Rightarrow\left({\cal B}\left[I-{\Pi_I\over I-\Pi_0}\right]{\cal
B}\right)
\end{eqnarray*}

Finally the binding energy is defined by the equation:
\begin{eqnarray}
\label{F-bind} &&1=\max\limits_U{\Pi_I[U]\over 1-\Pi_0[U]}
=\max\limits_V{(V\Pi_I V)\over\left(V \left[{1\over{\cal
D}}-\Pi_0\right]V\right)}= \max\limits_\Psi{(\Psi D\Pi_I
D\Psi)\over\left(\Psi \left[D-D\Pi_0D\right]\Psi\right)}.\nonumber\\
\end{eqnarray}

\subsection{Nonrelativistic limit}

Let us consider the nonrelativistic limit in the equation
(\ref{F-bind}) which takes place for small coupling constant. We
should take
$$\Psi({\bf y},y_4)=\Psi({\bf y})$$\
with
\begin{eqnarray*}
&&(\Psi D\Psi)={1\over4\pi}\int\! d{\bf y}~{\Psi^2({\bf
y})\over|{\bf y}|},
\end{eqnarray*}

Equation (\ref{F-bind}) becomes
\begin{eqnarray}
\label{F-bind-N} &&1=\max\limits_\Psi{(\Psi D\Pi_I D\Psi)\over(\Psi
D\Psi)}
\end{eqnarray}

For the polarization operator one can get
\begin{eqnarray*}
&&\tilde{\Pi}_I(q)=2e^2\int{d{\bf k}dk_4\over(2\pi)^4}V^2({\bf
k},k_4)~ {\left(m^2+{M^2\over4}\right)
\over \left({\bf k}^2+k_4^2+m^2-{M^2\over4}\right)^2+M^2k_4^2}\\
&&\Rightarrow 4e^2m^2\int{d{\bf k}dk_4\over(2\pi)^4}{V^2({\bf k})
\over \left({\bf k}^2+m\epsilon\right)^2+4m^4k_4^2} =e^2m \int{d{\bf
k}\over(2\pi)^3}{V^2({\bf k})\over {\bf k}^2+m\epsilon}
\end{eqnarray*}
In order to get the nonrelativic Schr\"{o}dinger equation, the term
$k^2_4$ in the denominator should be neglected (for details see
\cite{Ef-sc}).

Finally, we have the equation
\begin{eqnarray*}
&&1=\alpha m\max\limits_\Psi{\int\!\!\!\int d{\bf y}d{\bf y}'
~\Psi({\bf y}){1\over -\triangle+m\epsilon}\Psi({\bf y}')\over
\int\! d{\bf y}~{\Psi^2({\bf y})\over|{\bf y}|}},
\end{eqnarray*}
which is nothing else but the nonrelativistic Schr\"{o}dinger
equation.

In the non-relativistic case
\begin{eqnarray}\label{PsiNon}
&&\Psi({\bf y})=e^{-ay},
\end{eqnarray}
and we get the well known result
\begin{eqnarray*}
&&1=2\alpha
m~\max\limits_a{a\over(a+\sqrt{m\epsilon})^2}~~~\Longrightarrow~~~
\epsilon={\alpha^2\over4}m.
\end{eqnarray*}

\subsection{Variation calculations}

Let us come back to equation (\ref{F-bind}). For the lowest state of
orthopositronium the test function is chosen in the form
\begin{eqnarray}
\label{PsiVar} && \Psi({\bf y},y_4)=e^{-a\sqrt{{\bf y}^2+by_4^3}},
\end{eqnarray}
where $a$ and $b$ are variational parameters. This function is the
closest to the nonrelativistic wave function (see \cite{Ef-sc}).

Then the vertex function is
\begin{eqnarray}
\label{R} V(k)&=&\int dy~D(y)\Psi({\bf y},y_4)e^{-iky}=
\int{dy\over(2\pi)^2y^2}e^{-a\sqrt{{\bf y}^2+by_4^2}-i{\bf
ky}-ik_4y_4}
\nonumber\\
&=&{ab+\sqrt{a^2b+{\bf k}^2b+s^2}\over[{\bf
k}^2+s^2+a^2(1+b)+2a\sqrt{a^2b+{\bf k}^2b+s^2}]
~\sqrt{a^2b+{\bf k}^2b+s^2}}\nonumber\\
&=&R(k,s;a,b),~~~~~k=|{\bf k}|,~~~s=|k_4|;
\end{eqnarray}
and
\begin{eqnarray*}
(\Psi D\Psi)&=&\int{dy\over(2\pi)^2y^2}e^{-2a\sqrt{{\bf
y}^2+by_4^2}} ={1\over4a^2(1+\sqrt{b})}
\end{eqnarray*}
The positronium binding energy $\epsilon=2m-M$ is defined by the
equation
\begin{eqnarray}
\label{varmass} &&
1=2\alpha\left(1+{M^2\over4m^2}\right)\max\limits_{a,b}{8\over\pi^2}
\int\limits_0^\infty dk\int\limits_0^\infty dv\cdot
{k^2~a^2(1+\sqrt{b})R^2(k,s;a,b)\over
(k^2+s^2+\Delta)^2+4(1-\Delta)s^2}.\nonumber\\
&&\Delta=1-{M^2\over4m^2}={\epsilon\over
m}\left(1-{\epsilon\over4m}\right),~~~~~M=2m-\epsilon.
\end{eqnarray}

Since of $\Delta$ is small, it is convenient to introduce the new
variables
\begin{eqnarray*}
&& k\rightarrow \sqrt{\Delta}~k,~~~~s\rightarrow \Delta
s,~~~~a\rightarrow \sqrt{\Delta}~a,~~~~b\rightarrow \sqrt{\Delta}~b.
\end{eqnarray*}
We get
\begin{eqnarray*}
&&H(k,s,\Delta;a,b)\\
&&={\sqrt{\Delta}~ab+\sqrt{a^2b+{\bf k}^2b+s^2}\over[{\bf
k}^2+\Delta~s^2+a^2(1+\Delta b)+2a\sqrt{\Delta(a^2b+{\bf
k}^2b+s^2)}] ~\sqrt{a^2b+{\bf k}^2b+s^2}}\nonumber
\end{eqnarray*}
and our equation takes the form
\begin{eqnarray}\label{VarEq1}
&&1={\alpha\over\sqrt{\Delta}}\left(1+{M^2\over4m^2}\right)\max\limits_{a,b}{16\over\pi^2}
\int\limits_0^\infty dk\int\limits_0^\infty dv\cdot
{k^2~a^2(1+\sqrt{\Delta}~b)H^2(k,s,\Delta;a,b)\over
(k^2+\Delta~s^2+1)^2+4(1-\Delta)s^2}.\nonumber\\
\end{eqnarray}
Preliminary calculations have shown that the parameter $a$ is very
close to one and the parameter $b$ is very small for all coupling
constants $\alpha\leq0.5$, so that we can put $a=1$ and $b=0$ in the
limits of our calculation accuracy. Thus, the test function
practically coincides with the non-relativistic wave function
$\Psi(r)=e^{-\sqrt{\Delta}~r}$.

In the case $ a=1,~b=0$ the function $H$ is
\begin{eqnarray*}
&&H(k,s,\Delta;1,0)={1\over k^2+(1+\sqrt{\Delta}~s)^2}
\end{eqnarray*}

Finally, the equation defining binding energy of the
ortho-positronium looks like
\begin{eqnarray}\label{VarEq2}
&&1={\alpha\over\sqrt{\Delta}}\left(1-{\Delta\over2}\right){32\over\pi^2}
\int\limits_0^\infty dk\int\limits_0^\infty dv\cdot {k^2~
H^2(k,s,\Delta;1,0)\over (k^2+\Delta~s^2+1)^2+4(1-\Delta)s^2}.
\end{eqnarray}
One should stress that the difference between relativistic and
non-relativistic cases is defined by the term $k_4^2=\Delta~s^2$ in
the denominator of the fermion loop.

The results of numerical calculations are shown in Table 2.

For semiquantitative calculations one can use the approximation
\begin{eqnarray*}
&&{32\over\pi^2} \int\limits_0^\infty dk\int\limits_0^\infty dv\cdot
{k^2~ H^2(k,s,\Delta;1,0)\over
(k^2+\Delta~s^2+1)^2+4(1-\Delta)s^2}\approx{1\over2\sqrt{1+9\sqrt{\Delta}+6\Delta}}
\end{eqnarray*}
so that the equation become the form
\begin{eqnarray*}
&&{1\over\alpha}={2-\Delta\over4\sqrt{\Delta(1+9\sqrt{\Delta}+6\Delta})}~~~~{\rm
or}~~~~\alpha={4\sqrt{\Delta(1+9\sqrt{\Delta}+6\Delta})\over2-\Delta}.
\end{eqnarray*}
This formula gives semi-quantitative dependence of the binding
energy $\Delta$ on the coupling constant $\alpha$.

\section{Coulomb gauge}

Now let us consider the Coulomb gauge. In this case the Firz
transformations lead to
\begin{eqnarray*}
&&(\gamma_\mu)_{\alpha\beta}D_{\mu\nu}(y)(\gamma_\nu)_{\rho\sigma}\\
&&=\sum\limits_{J_1,J_2}(O_{J_1})_{\alpha\sigma}(O_{J_2})_{\rho\beta}
{1\over16}{\rm Tr}[O_{J_1}i\gamma_\mu
O_{J_2}i\gamma_\nu]D_{\mu\nu}(y)\\
&&\Longrightarrow-(\gamma_j)_{\alpha\sigma}{\cal
D}^V_{jj'}(y)(\gamma_{j'})_{\rho\beta}-
(i\gamma_0\gamma_j)_{\alpha\sigma}{\cal
D}^T_{jj'}(y)(i\gamma_0\gamma_{j'})_{\rho\beta} .
\end{eqnarray*}
where
\begin{eqnarray}\label{DCVT}
&&{\tilde{\cal D}}^V_{ij}(k)={1\over4}\left\{{\delta_{ij}\over{\bf
k}^2}-2{k_ik_j\over{\bf k}^2}\cdot{1\over k^2}\right\}~\Rightarrow~
{1\over4}{\delta_{ij}\over{\bf k}^2},\\
&&{\tilde{\cal D}}^T_{ij}(k)={1\over4}\left\{{\delta_{ij}\over{\bf
k}^2}+2{k_ik_j\over{\bf k}^2}\cdot{1\over k^2}\right\}~\Rightarrow~
{1\over4}{\delta_{ij}\over{\bf k}^2}.\nonumber
\end{eqnarray}
It means that in the Coulomb gauge the ortho-positronium is
described by a mixture of vector $V$ and tensor $T$ relativistic
currents.

In what follows we neglect the term ${k_ik_j\over{\bf k}^2}$. There
are two reasons to do it. First, usually in the generally accepted
approaches these terms are not considered at all. Second, we did not
have courage to perform these cumbersome calculations although they
can be done in case of emergency. Thus we have
\begin{eqnarray*}
&&\tilde{{\cal D}}_{ij}^V(k)=\tilde{{\cal D}}_{ij}^T(k)
=\tilde{{\cal D}}_{ij}(k)={1\over4}{\delta_{ij}\over {\bf k}^2}
\end{eqnarray*}
and
\begin{eqnarray*}
&&{\cal D}_{ij}(y)=\int{dk\over(2\pi)^4}\tilde{{\cal
D}}_{ij}(k)e^{iky}={\delta_{ij}\over16\pi}\cdot{\delta(y_4)\over|{\bf
y}|}
\end{eqnarray*}
The vector and tensor currents are
\begin{eqnarray*}
J^V_j(x,y)&=&(\overline{\Psi}(x_1)i\gamma_j\Psi(x_2))
=\left(\overline{\Psi}\left(x\right)e^{i{y\over2}\stackrel{\leftrightarrow}{p}_x}
\gamma_j\Psi\left(x\right)\right)\\
J^T_j(x,y)&=&(\overline{\Psi}(x_1)i\gamma_0\gamma_j\Psi(x_2))
=\left(\overline{\Psi}\left(x\right)e^{i{y\over2}\stackrel{\leftrightarrow}{p}_x}
i\gamma_0\gamma_j\Psi\left(x\right)\right)\\
&&~~~~~~~~~~~~\stackrel{\leftrightarrow}{p}_x= {1\over
i}\left(\stackrel{\leftarrow}{\partial}_x-
\stackrel{\rightarrow}{\partial}_x\right)
\end{eqnarray*}
The one-photon exchange term containing vector and tensor currents
looks like
\begin{eqnarray*}
&&W_2={e^2\over2}\int\!\!\!\int dxdy~\left[ J_\mu^V(x,y)\cdot{\cal
D}_{\mu\nu}^V(y)+J_\mu^T(x,y)\cdot{\cal D}_{\mu\nu}^V(y)
\cdot J_\nu^T(x,-y)\right]\\
&&={\alpha\over8}\int dx\int\!\!\!\int d{\bf y}_1d{\bf y}_2~\left[
{J_j^V(x,{\bf y}_1)\over\sqrt{|{\bf y}_1|}} \cdot\delta({\bf
y}_1-{\bf y}_2)\cdot{J_j^V(x,-{\bf y}_2)\over\sqrt{|{\bf
y}_2|}}\right.\\
&&~~~~~+\left. {J_j^T(x,{\bf y}_1)\over\sqrt{|{\bf y}_1|}}
\cdot\delta({\bf y}_1-{\bf y}_2)\cdot{J_j^T(x,-{\bf
y}_2)\over\sqrt{|{\bf y}_2|}}\right]
\end{eqnarray*}

In the Coulomb gauge we should introduce an orthonormal system in
the ${\bf x}\in{\bf R}^3$ space only:
\begin{eqnarray*}
&& \{U_Q({\bf y})\}=\{U_{nlm}({\bf y})\}=\left\{
\begin{array}{l}
\int d{\bf y}~U_{Q_1}^*({\bf y})U_{Q_2}({\bf y})=\delta_{Q_1,Q_2},\\
\\
\sum\limits_Q~U_Q({\bf y}_1)U_Q^*({\bf y}_2)=\delta({\bf y}_1-{\bf y}_2),\\
\end{array}\right.
\end{eqnarray*}

For ortho-positronium $Q=0$ and $U_0({\bf y})=U({\bf y})$ with
normalization $(UU)=1$.

We have
\begin{eqnarray*}
&&W_2~\Rightarrow~{\alpha\over8}\int dx\left[{\bf J}^V(x){\bf
J}^V(x)+{\bf J}^T(x){\bf J}^T(x)\right].
\end{eqnarray*}
Here
\begin{eqnarray*}
&&{\bf J}^V(x)= \left(\overline{\Psi}(x)
V(\stackrel{\leftrightarrow}{{\bf p}}_x)\mbox{\boldmath$\Gamma$}_V
\Psi(x)\right),~~~~ \mbox{\boldmath$\Gamma$}_V=\mbox{\boldmath$\gamma$}\\
&&{\bf J}^T(x)= \left(\overline{\Psi}(x)
V(\stackrel{\leftrightarrow}{{\bf p}}_x)\mbox{\boldmath$\Gamma$}_T
\Psi(x)\right),~~~~
\mbox{\boldmath$\Gamma$}_T=i\gamma_0\mbox{\boldmath$\gamma$}\\
\end{eqnarray*}
with
\begin{eqnarray*}
&&V({\stackrel{\leftrightarrow}{{\bf p}}_x})= \int d{\bf y}~ {U({\bf
y})\over\sqrt{|{\bf y}|}}e^{i{\bf y}{\stackrel{\leftrightarrow}{{\bf
p}}_x}}
\end{eqnarray*}

The generating functional takes the form
\begin{eqnarray}\label{ZZ}
Z&=&\int\!\!\! D\Psi D\overline{\Psi} e^{(\overline{\Psi}S^{-1}\Psi)
+{\alpha\over8}\int dx~[{\bf
J}_V^+(x){\bf J}_V(x)+{\bf J}_T^+(x){\bf J}_T(x)]}\\
&=&\int\!\!\! D\Psi D\overline{\Psi}e^{(\overline{\Psi}S^{-1}\Psi)}
\int\!\!\!\int D{\bf B} e^{-{1\over2}({\bf
BB})+{\sqrt{\alpha}\over2} (\overline{\Psi}{\cal B}\Psi)}\nonumber
\end{eqnarray}
where
\begin{eqnarray*}
&&D{\bf B}= D{\bf B}_VD{\bf B}_T,~~~~~({\bf BB})= ({\bf B}_V{\bf
B}_V) +({\bf B}_T{\bf B}_T)\\
&&{\cal B}=V\mbox{\boldmath$\Gamma$}_V{\bf B}_V
+V\mbox{\boldmath$\Gamma$}_T{\bf B}_T.
\end{eqnarray*}

The integration over electron variables gives
\begin{eqnarray*}
&&\int\!\!\!\int D\Psi D\overline{\Psi}
e^{(\overline{\Psi}S^{-1}\Psi)+{\sqrt{\alpha}\over2}
(\overline{\Psi}{\cal B}\Psi)}=\exp\left\{{\rm
Tr}\ln\left[1+{\sqrt{\alpha}\over2}
{\cal B}S\right]\right\}\nonumber\\
&\approx&e^{-{\alpha\over8}\int\!\!\!\int\! dx_1dx_2{\rm Tr} {\cal
B} S(x_1-x_2){\cal B} S(x_2-x_1)]}=e^{{1\over2}\int
{dp\over(2\pi)^4}( {\bf B}(p)\Pi(p^2){\bf B}(p))}.
\end{eqnarray*}
and the generating functional becomes the form
\begin{eqnarray}\label{ZZZ}
Z &=& \int\!\!\!\int D{\bf B} e^{-{1\over2}({\bf BB})+{1\over2}(
{\bf B}\Pi{\bf B})}
\end{eqnarray}

Here
\begin{eqnarray}
\label{DPP}
&&\Pi(p^2)= \left(\begin{array}{rr}
\Pi^{(VV)}(p^2), & \Pi^{(VT)}(p^2)\\
\Pi^{(TV)}(p^2), & \Pi^{(TT)}(p^2)\\
\end{array}\right),~~~~
{\bf B}(p)=\left(\begin{array}{c} {\bf B}_V(p)\\
{\bf B}_T(p)\\
\end{array}\right),\nonumber\\
&&\Pi^{(J_1J_2)}(p^2)=\alpha\int{dk\over(2\pi)^4}~{V^2({\bf
k})~T^{(J_1J_2)}_{ij}(k,p)
\over\left(m^2+\left(k+{p\over2}\right)^2\right)
\left(m^2+\left(k-{p\over2}\right)^2\right)}
\end{eqnarray}

\begin{eqnarray*}
T^{(J_1J_2)}_{ij}(k,p)&=&-{1\over4}{\rm
Tr}\left[\mbox{\boldmath$\Gamma$}_{J_1,i}\left(m+\hat{k}+{\hat{p}\over2}\right)
\mbox{\boldmath$\Gamma$}_{J_2,j}\left(m+\hat{k}-{\hat{p}\over2}\right)\right]
\end{eqnarray*}

Direct calculations give
\begin{eqnarray}
\label{TJJ}&&T_{ij}=\left(\begin{array}{cc}
T^{(VV)} &  T^{(VT)}\\
T^{(TV)} &  T^{(TT)} \\
\end{array}\right)_{ij}=\delta_{ij}(T_0+T_I),\\
&&T_0=\left(
\begin{array}{cc}
k_4^2+{{\bf k}^2\over3}& 0\\
0& k_4^2-{{\bf k}^2\over3}\\
\end{array}\right),~~~~~T_I= \left(
\begin{array}{cc}
m^2+{M^2\over4}&  i m M\\
-i m M& m^2+{M^2\over4}\\
\end{array}\right)\nonumber
\end{eqnarray}
The matrix $T_0$ is responsible for a continuous spectrum and should
be removed. The matrix $T_I$ is responsible for bound states and can
be represented as
\begin{eqnarray}
&&T_I= {1\over2}\left[\left(m+{M\over2}\right)^2
\left(\begin{array}{c}
1\\
-i \\
\end{array}\right)
\begin{array}{cc}
(1&  i) \\
~~&~~\\
\end{array}+
\left(m-{M\over2}\right)^2 \left(\begin{array}{c}
1\\
i \\
\end{array}\right)
\begin{array}{cc}
(1&  -i) \\
~~&~~\\
\end{array}\right]\nonumber
\end{eqnarray}

The eigenvalues are
\begin{eqnarray}
\label{MS}
&&\Lambda_\pm=\left(m\pm{M\over2}\right)^2
\end{eqnarray}
For the eigenvalue
$\Lambda=\left(m-{M\over2}\right)^2=m^2\left(1-{M\over2m}\right)^2
=m^2\left({\epsilon\over2m}\right)^2\ll m^2$ the bound state can
exist  but this case requires a very large coupling constant
$\alpha$ to provide the condition
$${\alpha\over\pi}{\Lambda\over
m^2}={\alpha\over\pi}\left({\epsilon\over2m}\right)^2\sim 1$$ This
condition cannot be realized (see \cite{Ef-frm}).

The most favorable situation for the existence of the positronium
takes place  for
$$\Lambda=\left(m+{M\over2}\right)^2=m^2\left(1+{M\over2m}\right)^2
\approx 4m^2$$ .

The quadratic form in the representation (\ref{ZZZ}) can be written
in the form
\begin{eqnarray*}
&&({\bf BB})-({\bf B}\Pi_I{\bf
B})=(W^+_+[1-\Pi_+]W_+)+(W^+_-[1-\Pi_+]W_-),\\
&&W_\pm={1\over\sqrt{2}}({\bf B}_V\pm i{\bf B}_T),\\
&&\Pi_\pm=\alpha\int{dk\over(2\pi)^4}~{V^2({\bf
k})\left(m\pm{M\over2}\right)^2
\over\left(m^2+\left(k+{p\over2}\right)^2\right)
\left(m^2+\left(k-{p\over2}\right)^2\right)}
\end{eqnarray*}
The combination $W_+={1\over\sqrt{2}}({\bf B}_V+i{\bf B}_T)$ with
$\Pi_+$ describes the desirable ortho-positronium $1^-$ state. The
Dirac spinor of this state equals
\begin{eqnarray*}
&&u_+=(1+\gamma_0)u=\left(\begin{array}{cc} 2 & 0\\
0 & 0 \\
\end{array}\right)\left(\begin{array}{c} 1\\
{\mbox{\boldmath$\sigma k$}\over m+E} \\
\end{array}\right)w=2\left(\begin{array}{c} 1\\
0\\
\end{array}\right)w
\end{eqnarray*}
according to a standard approach (see, for example, \cite{Berest}).

Finally, the partition function reads
\begin{eqnarray*}
&& Z_+=\int DW_+DW_+^+ e^{-{1\over 2}(W_+^+[1-\Pi_+]W_+)}
\end{eqnarray*}
The ortho-positronium bound state equation looks like
\begin{eqnarray}\label{OrthEq}
&& 1=\Pi_+
\end{eqnarray}

\subsection{Numerical calculations}

For numerical calculations let us choose the test function in the
form
\begin{eqnarray*}
&& U({\bf y})={\Psi({\bf y})\over\sqrt{|{\bf y}|}}={e^{-m a|{\bf
y}|}\over\sqrt{|{\bf y}|}},~~~~~||U||^2=(UU)={\pi\over m^2a^2}
\end{eqnarray*}
so that
\begin{eqnarray*}
&& V({\bf k})={1\over||U||} \int d{\bf y}~{U({\bf
y})\over\sqrt{|{\bf y}|}}e^{i({\bf ky})}={ma\over\sqrt{\pi}}\int
d{\bf y}~{e^{-m a|{\bf y}|}\over|{\bf y}|}e^{i({\bf
ky})}={4am\sqrt{\pi}\over {\bf k}^2+m^2a^2}.
\end{eqnarray*}
In our calculations we have only one variational parameter $a$.

After some calculations equation (\ref{OrthEq}) can be represented
in the form
\begin{eqnarray*}
&& 1=\max\limits_a\Pi_+=\max\limits_a{\alpha\over\sqrt{\Delta}}
\left(1+{M\over2m}\right)^2{2a^2\over\pi} \int\limits_0^\infty
{dv~v^2~ H\left(\Delta(v^2+1),{M^2\over m^2}\right) \over(v^2+1)(
v^2+a^2)^2}
\end{eqnarray*}
where
\begin{eqnarray*}
&& H(c^2,b^2)=\int\limits_{-\infty}^\infty
{dk_4\over(2\pi)}{4c^2\over
\left(k_4^2+c^2\right)^2+k_4^2b^2}\\
&&={2\sqrt{2}\over[\sqrt{2c^2+b^2+b\sqrt{4c^2+b^2}}+\sqrt{2c^2+b^2-b\sqrt{4c^2+b^2}}]}
\end{eqnarray*}
The results of the numerical calculations are given in Table 2. The
accuracy of these calculations is about $1\div2$ \% (see
\cite{Ef-sc}).

\vspace{1cm}

\newpage

\begin{center}
Table 2. Binding energy $\epsilon$ ($eV$) of the state $1^-$ as a
function of the coupling constant $\alpha$ for different gauges.

\vspace{.5cm}

\begin{tabular}{|c|c|c|c|c|c|c|c|}
\hline
&&&&&&&\\
$\alpha$&$0.0005$ & $0.001$&${1\over137}=0.0073$& 0.01 & 0.1& 0.3 & 0.5\\
&&&&&&&\\
\hline\hline
&&&&&&&\\
Feynman &0.032  & 0.126  & 6.47 & 12.0 & 893  & 5~700 & 12~600 \\
&&&&&&& \\
\hline
&&&&&&&\\
Coulomb&0.032&0.127& 6.8 & 12.8 & 1~270  & 10~800 & 27~800\\
&&&&&&&\\
\hline
&&&&&&&\\
Schr\"{o}dinger&0.032&0.127& 6.8  & 12.8  & 1~280  & 11~500 & 31~900 \\
${\alpha^2\over4}m_e$&&&&&&&\\
&&&&&&&\\
\hline
\end{tabular}
\end{center}

\section{Conclusion}

In conclusion, let us formulate our results.
\begin{itemize}
\item The Feynman and Coulomb gauges give coinciding results for
very small coupling constants $\alpha\leq0.1\alpha_{QED}$.
\item For $\alpha_{QED}={1\over 137}$ the difference of binding energies
for the Feynman and Coulomb gauges is of an order of $\sim 5\%$;
\item For $\alpha\sim0.1\div0.5$ difference in the binding energies
is of an order of $\sim 100\%$.
\item Calculations in the Bethe-Salpeter equation with the Coulomb gauge
and in the non-relativistic Schr\"{o}dinger equation coincides up to
$\alpha\leq0.1$.
\end{itemize}
Thus, one can conclude that in gauge theories like QCD, where the
coupling constant is not too small, the Bethe-Salpeter equation in
the one-"gluon" exchange approximation gives quite different numbers
for different gauges and, therefore, it is not a good mathematical
instrument for calculation of binding energies of bound states. One
can say that the gauge invariance is broken in the Bethe-Salpeter
equation with any fixed kernel. An alternative way is to recognize
that there exists a preferred gauge, namely the Coulomb gauge. This
idea is not new (see \cite{Perv} and other references there).

Besides, in real QCD we have an additional difficulty: the formation
of mesonic bound states takes place at large distances where
confinement plays the main role and we should know the explicit form
of quark and gluon propagators in the confinement region.

\vspace{.5cm}

The author is grateful to E.A.Kuraev and V.N.Pervushin for useful
and stimulating discussions.

\end{document}